# Who Re-Uses Data? A Bibliometric Analysis of Dataset Citations

Geoff Krause[1]*, Madelaine Hare[2], Mike Smit[3], Philippe Mongeon[4]
[1] gkrause@dal.ca
https://orcid.org/0000-0001-7943-5119
Department of Information Science, Faculty of Management, Dalhousie University, Canada
*Corresponding author

[2] maddie.hare@dal.ca
https://orcid.org/0000-0002-2123-9518
Department of Information Science, Faculty of Management, Dalhousie University, Canada

[3] mike.smit@dal.ca
https://orcid.org/0000-0002-2028-4317
Department of Information Science, Faculty of Management, Dalhousie University, Canada

[4] pmongeon@dal.ca
https://orcid.org/0000-0003-1021-059X
Department of Information Science, Faculty of Management, Dalhousie University, Canada
Centre interuniversitaire de recherche sur la science et la technologie (CIRST), Université du Québec à Montréal, Canada

## Abstract

Open data is receiving increased attention and support in academic environments, with one justification being that shared data may be re-used in further research. But what evidence exists for such re-use, and what is the relationship between the producers of shared datasets and researchers who use them? Using a sample of data citations from OpenAlex, this study investigates the relationship between creators and citers of datasets at the individual, institutional, and national levels. We find that the vast majority of datasets have no recorded citations, and that most cited datasets only have a single citation. Rates of self-citation by individuals and institutions tend towards the low end of previous findings and vary widely across disciplines. At the country level, the United States is by far the most prominent exporter of re-used datasets, while importation is more evenly distributed. Understanding where and how the sharing of data between researchers, institutions, and countries takes place is essential to developing open research practices.

## Introduction

Sharing data is an important pillar of Open Science and has commonly recognized benefits for science by reducing error, signaling credibility, and aligning with institutional or publisher mandates, but also for individual researchers who see greater visibility and citations of their work (Gleditsch et al., 2003; Drachen et al., 2016; Park & Wolfram, 2017; Zhang & Ma, 2023) and increased collaboration (McKiernan et al., 2016). The FAIR principles, designed to optimize the re-use of data, have been established to forward two major reasons for data re-use: the efficiency of knowledge production through effective sharing and the transparency of the scientific process.

Researchers may not be inclined to share their data for various reasons, such as the effort expenditure required (Parsons et al., 2019), the loss of publication opportunity, lack of credit, lack of time, fear of criticism, and data sensitivity. Even when data is shared, it may not be easy to find or to use by other researchers due to a lack of standardization in data formats and metadata and the multiplicity of platforms, for instance. Moreover, available data is not always re-usable or perceived as re-usable by other researchers. Potential re-users may lack confidence in the relevance, quality, or provenance of the data.

Data sharing practices and accompanying infrastructures on which they depend (Sá & Grieco, 2016) have been long-established in certain fields, such as economics, but are relatively nascent in others. Data itself is increasingly recognized as research output, and like other research outputs, there have been many efforts to develop ways of capturing their contributions to the advancement of knowledge. Infrastructures have been developed to support data-related endeavors, such as repositories where researchers can share their data (e.g., FigShare, Zenodo, The DataVerse), others that track citations to these datasets (e.g., the Data Citation Index and DataCite), or platforms aimed at enhancing the search and discovery of datasets (e.g., Elsevier DataSearch). The practice of data citation is facilitated by these infrastructures and has become an increasingly common mechanism for attributing and tracking dataset use (Borgman, 2016; Peters et al., 2016; Robinson-García et al., 2016). As Parsons et al. (2019) put it, data citation helps make data sharing both FAIR and fair- that is, increasing the efficiency of the scientific system and its transparency.

Past studies investigating data re-use found that formal data citations are generally rare (Park et al., 2018; Park & Wolfram, 2017; Peters et al., 2016; Robinson-García et al., 2016) and observed differences in data citation practices across disciplines (Peters et al., 2016). Park et al. (2018) found that the informal citation of data in the text of articles is more frequent than formal citations in reference lists, limiting documented credit for data contributions. These studies highlight the lack of standardized methods of signaling data re-use as an ongoing challenge to measuring the scholarly impact of data.

Wang et al. (2021) observed a lack of emphasis in past studies on the influence of the context of data re-use. This study fills this gap by providing a global overview of data citation practices that considers the data citation context. Given the large investments that the scientific community and other stakeholders have committed to data-sharing and re-use, it seems reasonable to question whether those investments are showing returns by investigating how much of the data is being used and by whom. The challenges of sharing, finding, or re-using data might lead one to hypothesize that when re-use does happen, it may be by researchers who have some existing familiarity with the data, starting with the data creators themselves.

Dudek et al. (2019) analyzed citations to datasets from the Institut Français de Recherche pour l'Exploitation de la Mer (IFREMER) and found that most of those citations came from researchers affiliated with the organization, and that 75% of the citing papers had at least one author in common with the dataset. Our work aims to scale up this work by using all datasets and their citation instances in OpenAlex, an open database that indexes over 240 million works. More specifically, we ask:

> **RQ1** How are citations to datasets in OpenAlex distributed, and what proportion are individual, institutional, and country self-citations?
> **RQ2** What sort of patterns emerge in the citing of datasets across different institutions or countries?
> **RQ3** How do the patterns of citations and re-use differ across disciplines?

This is, to our knowledge, the first analysis to provide large-scale empirical insights on the relationship between the data producers and the data re-users, shedding light on the social dynamics of data citation and re-use. Indeed, this study allows us to reflect on the effectiveness of data sharing practices and infrastructures in supporting one of the core narratives around open data: its ability to be efficiently re-used by other researchers to generate new knowledge. Moreover, this study allows us to question a “monolithic” perspective on data sharing and re-use by suggesting that if the moral argument for data sharing in the name of transparency and reproducibility may be valid across disciplines, the efficiency for knowledge production argument may be at the very least overstated.

# Literature review

In this section we review the literature on data re-use and data citations as concepts and as practices.

### Enablers and barriers of research data sharing

Sharing data is important for verifying and replicating published research, making findings available to the public, gaining new insights, and driving innovation in research (Borgman, 2012). It also leads to cost savings, improved collaboration, enhanced education, and better data (Uhlir & Schröder, 2007). Most researchers recognize the benefits of open data, yet open data practices are not widely adopted (Fecher et al., 2017; Tenopir et al., 2020). This may be explained by the many barriers to data sharing highlighted in the literature. These include fear of hindering one's career (Langat et al., 2011), time and effort required (Enke et al., 2012; Kim & Stanton, 2016; Kim & Zhang, 2015; Vickers, 2006; Zuiderwijk et al., 2020), fear of misuse (Enke et al., 2012), lack of resources (Campbell et al., 2002), concerns about scientific priority (Campbell et al., 2002), lack of institutional support for data management (Tenopir et al., 2011), lack of standards (Enke et al., 2012), and lack of acknowledgement and reward for data sharing (Belter, 2014; Enke et al., 2012; Molloy, 2011; Schäfer et al., 2011).

Data sharing may be better facilitated by the robust alignment of data sharing rationales, reward and incentive structures, policy, and technical components that support their operationalization (Borgman, 2016; Sá & Grieco, 2016). Data sharing can be incentivized through funding or journal policies that require data sharing (Borgman, 2012; Zuiderwijk et al., 2020), and advocacy and social influence (Zuiderwijk et al., 2020). Past experience of data re-use was also found to positively influence intent to share data (A. Yoon & Kim, 2020), as was a sense that data being collected might in some way be of use or interest to others (Wallis et al., 2013). Technological advancements can support open science by reducing barriers to storing and sharing data (Haak et al., 2020; Molloy, 2011; Sá & Grieco, 2016). Peer-reviewed "data papers" can also contribute to both the sharing and reward aspects by reframing data production and citation within the traditional scholarly communication process (Huang & Jeng, 2022; Molloy, 2011; Sá & Grieco, 2016). The development of open data metrics and other reward mechanisms like open data badges can also provide additional rewards and thus incentivize data sharing (Curty et al., 2017; Dorta-González et al., 2021; Lowenberg et al., 2019; Mongeon et al., 2017; Rowhani-Farid et al., 2017). Hood and Sutherland (2021), for example, advocate for a new author-level metric called the data index, calculated the same way as the h-index.

Another element that may incentivize data sharing is the data citation advantage for papers with an accompanying dataset (Drachen et al., 2016; Gleditsch et al., 2003; Piwowar & Vision, 2013). While a study by Zhang and Ma (2023) suggests that the citation advantage might be short-lived, dissipating after 4-5 years, this may not actually inhibit the motivational effect of a perceived potential for more citations associated with data sharing. He & Han (2017) found positive correlations between usage counts of datasets and citation counts of articles. In contrast, Thelwall and Kousha (2017) found only a weak relationship between downloads and citations.

## Data re-use and data citations

### Conceptual challenges

The dynamism of data affects the conceptualization of what constitutes a dataset, its tracking, and its attribution. Formal data citation, allowing for the use of data metrics for the purpose of academic performance evaluation (as one example) offers the potential to promote increased data re-use through engagement with the rewards system of science. Data citation is the endpoint of the process of discovering the data, using the data, and attributing credit to the data producers (Borgman, 2016). However, it is only once the resulting work is published that the re-use becomes public, and only once the work is added to a citation index that the re-use can be tracked and converted to a data metric. Furthermore, the different mediums through which data may be published, how it may arrive at the citation stage, and the citation paradigm have been argued to be incompatible with the needs of data reporting and management (Buneman et al., 2021; Zwölf et al., 2019). For instance, the fact that not all databases or datasets are static poses some conceptual and technical challenges for data citations which, according to Buneman et al. (2021), need to enable linking to different versions or sub-components of datasets. Federer (2020) synthesized recommendations for measuring and tracking data that highlight the challenges of defining the term "data" and determining whether different versions of datasets should be considered as separate entities. Problems with the lack of standardization of data citation across disciplines, including metadata standards, were identified as inhibitors to existing automated methods for tracking and quantifying them (Federer, 2020). Indeed, many articles point to the availability of the data but do not include a DOI or instructions on how to cite the data, which may represent a barrier to formal data citations as a measure of re-use (Zhao et al., 2018). Attempts have been made to remedy problems with data identification and attribution to incentivize data sharing:  considering the lack of widely accepted standards for data citation, Honor et al. (2016) present a prototype of a system that integrates DOIs and a standardized metadata scheme into repository workflows for the field of neuroimaging. Permanent identifiers such as DOIs undoubtably improve the unique identification of datasets, however Mayernik et al. (2017) note that it is not clear that this has actually increased citation or re-use.

Further, defining data poses challenges in distinguishing use from re-use, and in what researchers *do* with data once they have it. Mayo et al. (2016) note that distinguishing original data use from re-use through citations alone may prove difficult, as standards and policies are less clear on citing data in the original case. In practice, citing data for primary use has not been common (Mooney & Newton, 2012). Pasquetto et al. (2019) characterize data re-use as a broad concept comprising many activities; revisiting one's own data for new research, using publicly available datasets to compare with new data, conducting an environmental scan of available datasets for new research projects, and undertaking reanalyses to answer new questions all implicate the contextual and temporal relationships of data use and re-use. They define data re-use as usage of a dataset by someone other than the creator (Pasquetto et al., 2019). For the purposes of this analysis, however, we consider data use by the creator as re-use and identify this through self-citation. Investigating data re-use in two scientific consortia, Pasquetto et al. (2019) found that it occurred mainly for reproducibility, purposes of comparison, calibration and identifying baseline measurements, and integration in new analyses (Pasquetto et al., 2017, 2019). Sakai et al. (2020) found the most common form of data re-use to be the integration of existing data with new, original data (20.7%), though data was mainly sourced from other articles rather than data archives. Lafia et al. (2023) discuss the different ways in which researchers engage with datasets in the academic text, such as describing, contextualizing, comparing datasets, and reify that not all mentions of datasets indicate re-use or are in the form of a formal citation.

**Determinants of data re-use**

Various factors influence researchers' decisions to re-use data. Hemphill et al. (2022) analyzed data usage for 380 studies released by the Inter-university Consortium for Political and Social Research (ICPSR) and found that data attributes, curatorial decisions, and archive funding models correlated with data re-use, highlighting the importance of usability for re-use. The field of astronomy, for example, produces large-scale datasets, making them difficult to homogenize for information retrieval purposes, subsequently impacting their discovery, location, retrieval, storage, and eventual re-use (Sands et al., 2012). Many researchers are more inclined to create their own datasets rather than to re-use data (Zhao et al., 2018), potentially resulting from familiarity with its context, greater trust in its quality, and its accessibility.

In contrast, surveys of researchers across disciplines (Kim & Yoon, 2017) and in social sciences (A. Yoon & Kim, 2017) found that perceived usefulness is the main driver of data re-use, alongside access to institutional resources and disciplinary data repositories as significant factors. The effect of perceived effort was not statistically significant in their model, suggesting that researchers are willing to put in the effort to re-use data when it meets their needs. Similarly, Imker et al. (2021) found in their study of data re-use practices of faculty researchers that although data is often not re-used because it is not *usable*, researchers also do not re-use data because it may not be truly *useful* to their work. They found that irrelevance of data to the context of future studies, and the educational value in collection itself were major factors in data re-use. Sharing data for the purpose of re-use is fundamentally different, and more challenging, than sharing data for reproducibility purposes (Borgman et al., 2021). It seems plausible to hypothesize that the sharing of re-usable data is thus less frequent than simply making one's data available alongside a published paper for transparency purposes, or to satisfy a funder's requirement or social norms. This would likely lead to a concentration of re-use where a minority of available datasets are cited, and an even smaller minority receive the majority of the citations.

In this way, researchers' perceptions and attitudes towards data re-use are an influential factor in their decision to do so. Many researchers possess different outlooks toward sharing their own data than they do accessing others' data (Uhlir & Schröder, 2007). Some of the reasons why researchers may be either unable to find useful data or are unwilling to use it include a lack of trust in the quality of the data, a lack of metadata standards (A. B. Ninkov et al., 2022; Tenopir et al., 2020; A. Yoon & Lee, 2019), or accompanying contextual information (or its absence) (Jiao & Darch, 2020). Further, Curty (2016) found that social scientists' willingness to re-use data is influenced by their peers' opinion towards data re-use, particularly for novice researchers (Faniel et al., 2012). Their understanding of the original context of the data is also a factor and is facilitated by the ability to contact the original data creators or obtain training or support (Curty, 2016), the tacit knowledge gained through disciplinary training and data-gathering experience (Zimmerman, 2007; Imker et al., 2021). Data re-use can be facilitated by organizational support for contending with larger challenges such as ethical dilemmas like privacy, and navigating modern legal frameworks (Safran, 2017).

**Disciplinary differences**

As Gitelman and Jackson (2013) note, "one productive way to think about data is to ask how different disciplines conceive their objects, or, better, how disciplines and their objects are mutually conceived" (pg. 7). Indeed, differences in data types, formats, modes of sharing, and utility are all influential on how data is shared, cited, and re-used across disciplines and how researchers perceive usability and usefulness of data. Disciplinary differences in data needs, types of data used, and data-sharing and re-use perspectives and norms should be considered when attempting to raise researchers' awareness and to develop data sharing and re-use policies that are effective across disciplines (Khan et al., 2023). Gregory et al. (2023) explored data citation practices, preferences, and motivations in a large-scale survey-based study. They found that humanities researchers use qualitative data more than any other disciplinary group, and also tend to cite their own data more frequently than other groups (Gregory et al., 2023). Joo et al. (2017) found that perceived usefulness was the strongest indicator influencing data re-use in the health sciences while the attitudes of STEM scientists towards data re-use were largely influenced by metadata standards, data re-use norms, and repository availability (Kim, 2021). Earth sciences tend to use hard to reproduce observational data that possesses value beyond its original purpose, whereas chemistry and physics typically make use of reproducible experimental data (Jiao & Darch, 2020). Disciplines such as archaeology may place higher cultural value on fieldwork, making it more difficult to justify diverting resources to "desk" work needed to understand existing datasets available for re-use, or preparing one's own data for sharing (Sobotkova, 2018).

Overall, the heterogeneity of data and of the epistemic cultures that convert them into knowledge lead to varying degrees of data re-usability (Carlson & Anderson, 2007). Data that is numeric, standardized, and born digital may be easily shared and re-used whereas non-digital data (e.g., texts, other media, objects) require transformation to be shared and may lose some of their authenticity (and therefore value or trustworthiness) in the process (Carlson & Anderson, 2007). Carlson and Anderson (2007), argue that knowledge is often not easily separated from its producer and its context to be re-used by other researchers in different settings, and that these challenges exist for both qualitative and quantitative data. As Wang et al. (2021) illustrate, the factors influencing data re-use do not always act independently but are often interwoven.

## Data metrics

Data citations are necessary for the development of data-level metrics which are essential tools for recognition in our metrics-centered evaluation regimes (Cousijn et al., 2019). They exist alongside other data metrics such as dataset views and downloads (Lowenberg, 2022) that do not necessarily indicate the use of data but may still be useful for highlighting engagement with data beyond the confines of the scholarly communications cycle where most formal data citations occur (Lowenberg et al., 2019). Examples of data metrics include the Data-index (Hood and Sutherland (2021)) and DataRank (2020). The latter considers the fact that informal data citation still tends to be far more frequent than formal citations (Park et al., 2018), which are relatively uncommon in most fields (Peters et al., 2016; Robinson-García et al., 2016), impacting data metrics and highlighting discrepancies across disciplines. New bibliometric data sources for developing data metrics, such as DataCite, have been the focus of recent studies (Ninkov et al., 2021; Robinson-Garcia et al., 2017; Rueda et al., 2017) offering richer data than traditional sources like the Data Citation Index, however, both studies encountered challenges regarding what data is, disciplinary classification, and lack of metadata standardization, which impede researchers' abilities to fully harness this source for bibliometric analyses. Ultimately, even if conceptual challenges surrounding data citation are solved, citations that cannot be tracked reliably and exhaustively will prevent the development of effective metrics to measure data re-use (Federer, 2020; Stuart, 2017).

## Past bibliometric studies on data re-use and data citations

Several bibliometric studies investigated data citation and re-use. Mooney's (2011) study of social sciences articles from the early 2000s found that only 40% of articles performing secondary analysis of data provided any sort of reference or mention of the datasets. A follow-up study of works published in 2010 found that the majority referred to the data in-text rather than in works cited sections, and that most did so by title alone (Mooney & Newton, 2012). Yoon et al. (2019) found that most studies re-using data from Health Information National Trends Survey (HINTS) cited the related papers rather than the datasets. Bai and Du (2022) analyzed citations to clinical datasets using the Data Citation Index and found that only 7.5% of datasets were formally cited. However, data papers and data repositories, which amounted to less than 15% of the records analyzed, were more often cited with rates of 24.2% and 85%, respectively. Khan et al. 2021) investigated data citation and re-use practices by analyzing metadata of 43,802 datasets indexed in the Global Biodiversity Information Facility (GBIF) and performed content analysis on articles citing GBIF data. They found that open data on GBIF is frequently re-used, and a steady increase in the number of articles re-using and citing this data. However, Khan et al. (2021) also found that GBIF uses a semi-automated system that produces unique DOI and accession dates for data subsets found through search queries, which are referencing the original dataset leading to an inflation of dataset citations. Using Datacite, Ninkov et al. (2021) found that the majority of datasets (86%) were published in the last 10 years, and over 50% were in English. They found that of the 6% of the dataset with a disciplinary classification, most came from biological, earth and related environmental, and health sciences. These studies illustrate the existing work undertaken to understand data citation and re-use from a variety of angles; the presence of data sources, their engagement and use, and what they reveal about fields and disciplines all contribute to a larger picture of the current state of data re-use in the scientific landscape.

# Data and methods

This study uses a snapshot of OpenAlex (Priem et al., 2022), circa May 2022, hosted in a PostgreSQL server. This snapshot contains records for over 211 million works, out of which 531,299 are dataset records. As our work is focused on comparing re-use by individual authors, institutions, and countries, our focus is on those datasets with authorship information present in OpenAlex, therefore excluding 124,541 datasets with no authorship data on record. An additional 117,195 datasets, more than a quarter of those available, are attributed to a single author, "A. Boyle", at the University of Michigan–Ann Arbor; as these likely represent an artefact of OpenAlex's aggregation of data sources, and no citations to these are present in our dataset, we have excluded them from further analysis. This leaves us with a sample of 289,481 datasets.

Authors, institutions, and countries were retrieved for these datasets and for the works citing the datasets. Overall, a set of 112,577 citations between 106,026 citing works and 12,963 cited datasets was obtained after filtering out 11,533 citations to datasets with no authorship data and 327 citing works with no authorship data.

*Disciplines*

Citations to datasets were assigned to the "domain" level of the Science Metrix classification scheme (Archambault et al., 2011) by using the ISSN or journal name. Journals with a broad focus assigned to the "article-level classification" domain in the Science Metrix classification were tagged as unclassified for our analysis. Within our dataset, 79,348 citing works (about 75%), were published in one of 10,610 journals in the Science-Metrix classification. Of these, 2,906 were unclassified.

The remaining unclassified works (both those that were not assigned to a classified journal, and those in a journal assigned to the "article level classification") were then assigned classifications on the basis of their citations to other works, using the classifications of the journals in which these referenced works were published. Works were assigned to the domain classifications they most frequently referenced. This process resulted in an additional 9,791 citing works being classified, for a total of 86,233 distinct citing works making 91,862 citations to 11,913 distinct datasets.

Note that journals or works (and thus, citations) may be associated with multiple domains (including the case of ties in classification based on references), so counts of citations and works for domains will add up to more than 100%. This is more pronounced for counts of associated datasets, which may be referenced by works in multiple disciplines.

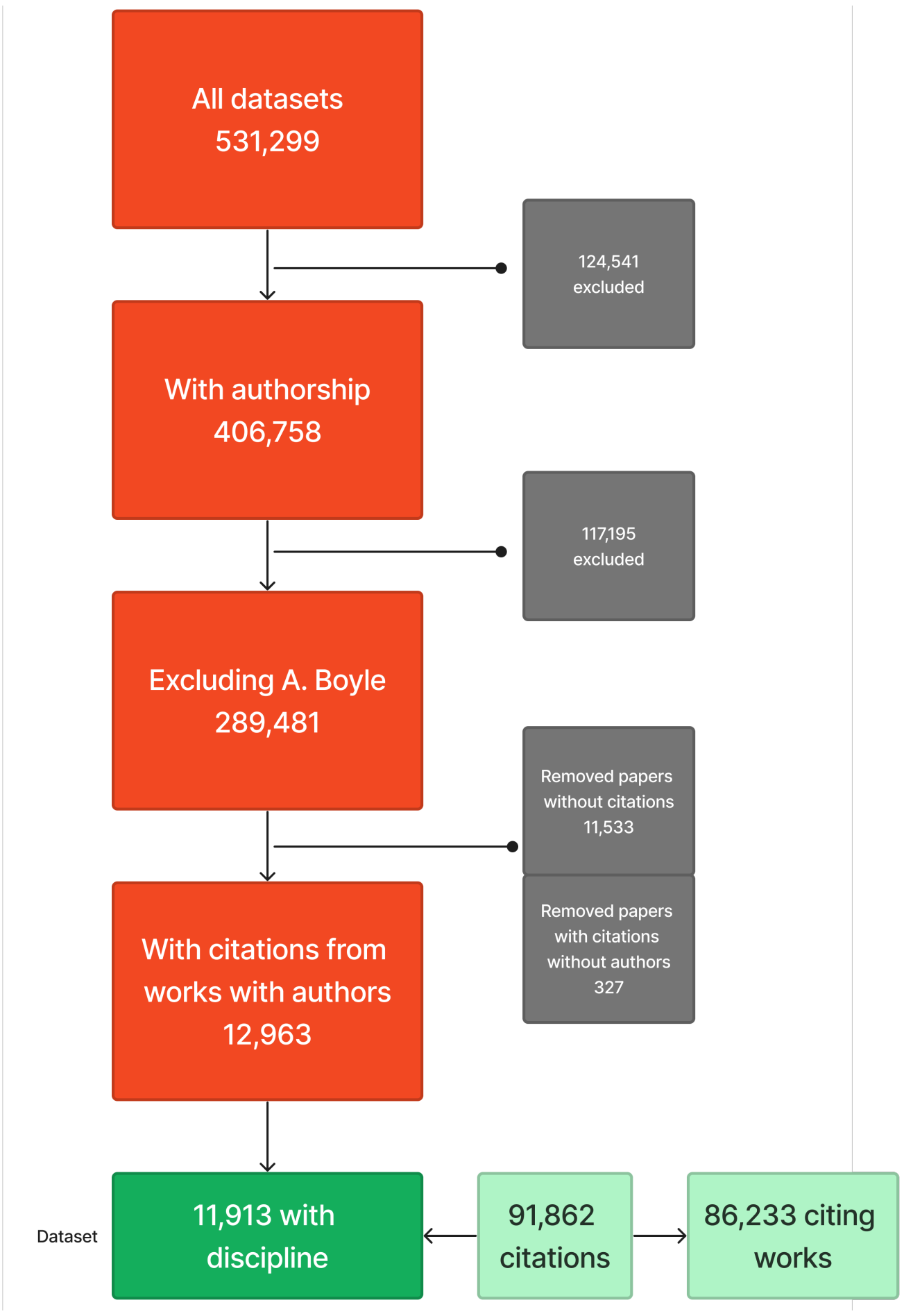

Figure 1. Dataset creation stages flowchart

*Institutional affiliations and country data*

The completeness of institutional and country authorship information is dependent on the metadata provided by the data sources used by OpenAlex. This information was available for at least one author of just under 90% of the citing works and 84% of all authorship records. However, fewer than 12% of all datasets have this information for any authors (around 7% of dataset-linked authorships), and 3.6% of the cited datasets (2.4% of authorships) with discipline classifications.

To address this issue, the works associated to the Author ID of any author or creator without a recorded affiliation were retrieved for a period of 2 years prior to 2 years after the publication of the dataset or citing work. The missing affiliation data was completed using the institutional affiliation listed on the work that is chronologically closest to the dataset or citing work. When no affiliation was found among the works within the 5-year window surrounding the work, the institution recorded in the "last_known_institution" field of the OpenAlex author record was used. This process resulted in institutional data being available for 68% of cited datasets and 97% of citing works.

*Identifying self-citations*

Self-citations were identified by comparing all authors on the dataset and the citing work. This includes matches across the OpenAlex Author ID, ORCID, exact display names from OpenAlex author records, and tokenized name groups. The latter consists of a series of variant name patterns created by a transformation to the Latin-ASCII character set, separating tokens into first/last name groups based on space separation (effectively creating a new variant for each space) and identifying both the first token and a set of initials for the "first name" grouping. Matches are then produced by comparing the "last name" groups, and then one or more of the full "first name" group, the "first name" initials, or the first token.

One limitation of this approach is that it may increase false positives in cases where the authors citing the dataset have a very common name. However, this is mitigated by the requirement of a single author-creator match for the citation to be considered a self-citation, meaning that cases of multiple matches between citing authors and data creators are more likely to be true positive, but also that the ambiguity of a common name is unlikely to impact our findings when a stronger match exists in the article and dataset bylines. The OpenAlex author ID could not be used alone because of a large degree of fragmentation of individual authors in the database, especially for non-article outputs that are difficult to link with other documents due to their systematically less complete metadata. For example, the previously mentioned "A. Boyle" datasets were linked to over 100,000 different author IDs.

*Final dataset*

Because of the significant number of observations for which we do not have affiliations data for either the dataset or the citing work, the parts of the analysis that focus on institutions and countries are limited to a subset of the full dataset. Table 1 presents the number of citations, distinct works, and distinct datasets included in the full dataset as well as the data subset for which institution data is available.

Table 1. Summary of datasets by venue, classification, and disciplinary characteristics

| | Full dataset | | | with institution data | | |
|---|---|---|---|---|---|---|
| | **Citations** | **Works** | **Datasets** | **Citations** | **Works** | **Datasets** |
| Applied Sciences | 7,881 | 7,530 | 2,437 | 5,714 | 7,368 | 1,649 |
| Arts & Humanities | 1,561 | 1,518 | 593 | 1,082 | 1,347 | 374 |
| Economic & Social Sciences | 13,725 | 13,043 | 2,653 | 10,079 | 12,419 | 1,821 |
| Health Sciences | 56,537 | 53,513 | 5,134 | 45,391 | 52,333 | 3,479 |
| Natural Sciences | 12,595 | 11,046 | 4,491 | 9,200 | 10,897 | 3,196 |
| Total | 91,862 | 86,233 | 11,913 | 71,189 | 83,986 | 8,164 |

*Analysis*

This paper focuses on the data producers, the data (re)-users, and the relationship between them, and thus requires that we establish indicators to measure data production and data usage. Data production is operationalized as a published dataset, and the producers as the individuals and by extension the institutions and countries listed as creators of the dataset. Data re-use is operationalized as the citation of a dataset in a work, the data re-users being any individual, and by extension the institutions and countries, listed as authors of that work. Matches between the two sets are counted to measure the total data outputs and uses and to analyze the distribution of data outputs and data re-uses across entities (individuals, institutions, and countries).

Three levels of self-citations are used to capture the relationship between the data re-users and creators: individual self-citations (at least one author or the citing paper is also listed as a creator of the dataset), institutional self-citations (the citing work and dataset share at least one institution), and country self-citations (the citing work and dataset share at least one country).Table 2 presents the variables that are used to perform the analyses.

Table 2. List of variables used for the analysis.

| **Variable** | **Description** |
|---|---|
| citing_work_id | OpenAlex Work ID (numeric portion only) of the citing work. |
| dataset_work_id | OpenAlex Work ID (numeric portion only) of the dataset being cited. |
| domains | Domain(s) from the Science Metrix classification associated to the citing work. |
| citing_institutions | OpenAlex Institution IDs (numeric portions only) associated with authorship of the citing work. Distinct values are separated by semi-colons. |
| citing_countries | Two-letter country codes associated with the institutions in the *citing_institution_list*. Distinct values are separated by semi-colons. |
| dataset_institutions | OpenAlex Institution IDs (numeric portions only) associated with authorship of the cited dataset. Distinct values are separated by semi-colons. |
| dataset_countries | Two-letter country codes associated with the institutions in the *dataset_institution_list*. Distinct values are separated by semi-colons. |
| author_match | Dichotomous variable indicating an author match between the citing work and the dataset. |
| institution_match | Dichotomous variable indicating an institution match between the citing work and the dataset. |
| country_match | Dichotomous variable indicating a country match between the citing work and the dataset. |

# Results

## Distribution of citations

The citation data from OpenAlex shows 86,233 works citing 12,963 datasets (4.5% of all datasets being examined), or around 7.7 citations per dataset.

Table 3 presents a summary of the number of citations by dataset for each discipline. The Health Sciences possess the highest citation rates with a mean of 11.0 citations per dataset and 2,292 as its maximum number of citations (cumulative citations, 45,391), though its standard deviation (72.3) is highest amongst all disciplines. The Health Sciences possesses the largest number of datasets amongst disciplines, with 643 more than the Natural Sciences. Despite having the second highest number of datasets (4,491) the Natural Sciences have the second lowest mean number of citations (2.80) of all disciplines. Applied Sciences, Natural Sciences, and Arts & Humanities, in descending order, receive the fewest mean citations, with 3.23, 2.80, and 2.63, respectively. Note that this is based on the classification of the citing articles – some datasets may see use in multiple disciplines, or in works that are published in journals that are not classified or are not limited to particular disciplines.

Table 3. Statistical summary of the number of citations by datasets for each discipline.

| Citing discipline | Number of Datasets | Number of citations | | | | |
|---|---|---|---|---|---|---|
| | | Mean | SD | Min | Median | Max |
| Applied Sciences | 2,437 | 3.23 | 23.1 | 1 | 1 | 867 |
| Arts & Humanities | 593 | 2.63 | 12.0 | 1 | 1 | 278 |
| Economic & Social Sciences | 2,653 | 5.17 | 24.0 | 1 | 1 | 654 |
| Health Sciences | 5,134 | 11.0 | 72.3 | 1 | 1 | 2,292 |
| Natural Sciences | 4,491 | 2.80 | 11.6 | 1 | 1 | 361 |
| Overall | 11,913 | 7.71 | 59.6 | 1 | 1 | 2,407 |

Nearly 58% of these datasets, 7,503, are cited only a single time, and these single citations account for less than 6.7% of all citation pairs. The most-cited dataset, the Beck Depression Index, has 2,407 citations, around 2.6% of all citation pairs, in itself. The top 10% of cited datasets account for 80% of citations. Figure 2 demonstrates the extreme skewness of citations across datasets.

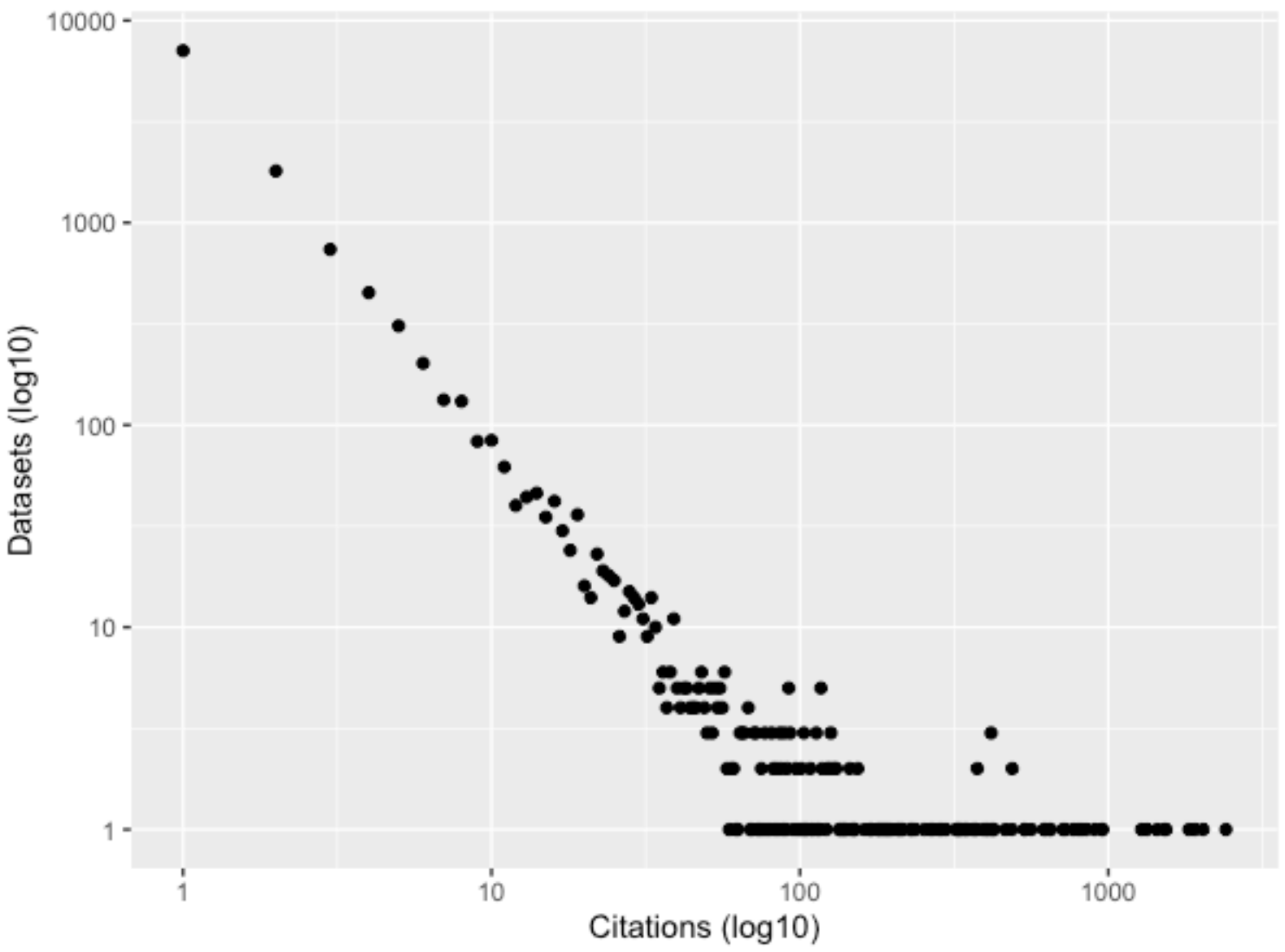

Figure 2. Datasets by number of citations

## Dataset producers

Institution data is available for 8,164 datasets, representing 4,745 different institutions in 145 countries. Figure 3 shows the top ten institutions, by number of datasets attributed to creators linked to them, for individual disciplines and overall.

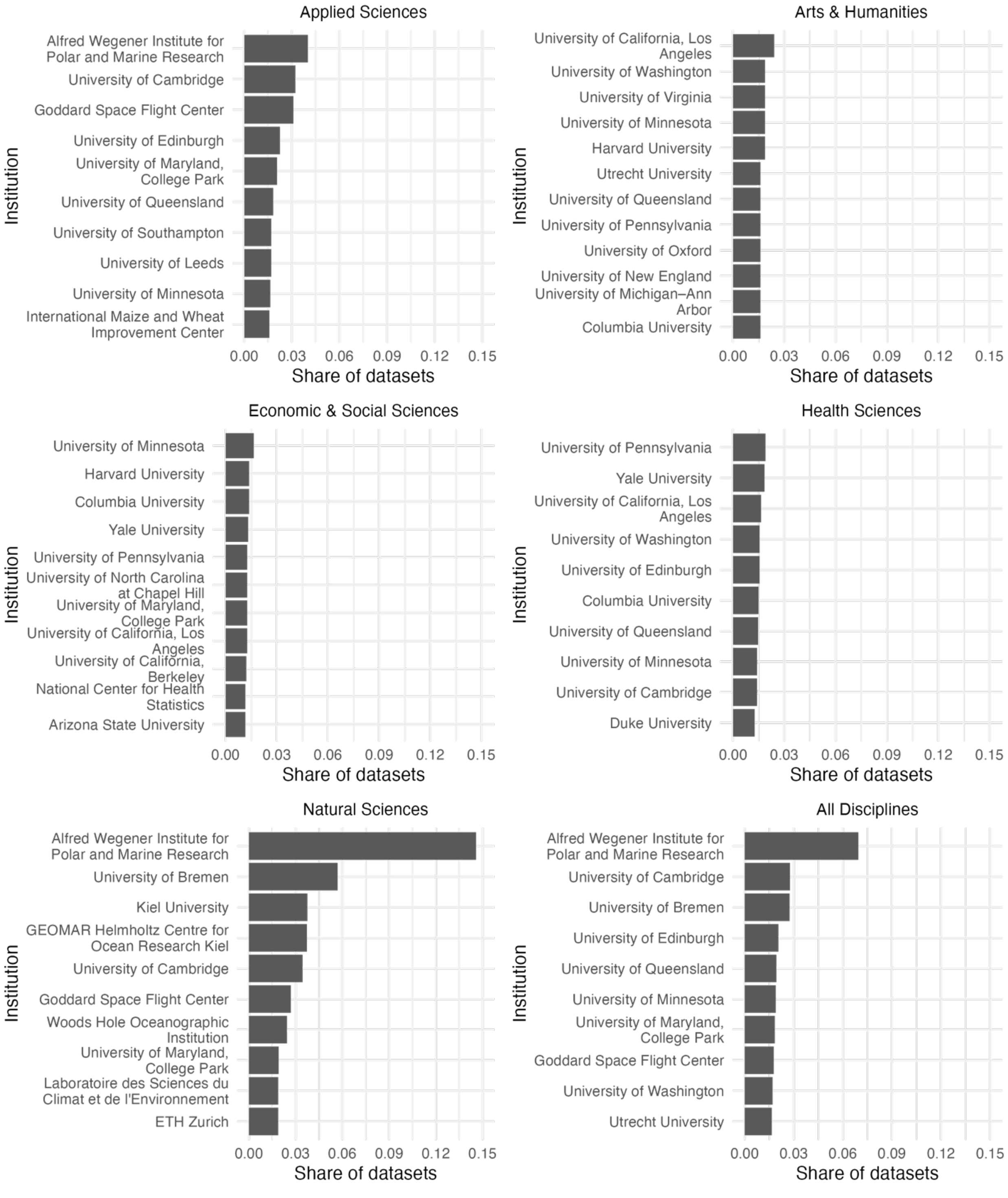

Figure 3. Top 10 dataset producing institutions by discipline

Though the United States holds a dominant position in the overall production of datasets and in those cited by other works, other countries, particularly the UK and Germany, place higher at the level of individual institutions, indicating a higher concentration of dataset production within specific institutions in these countries. This is particularly noticeable in the Natural and Applied Sciences, both at the level of the individual institutions and at the country level. Figure 4 shows the top ten countries for production of datasets, based on the location of institutions linked to the datasets' creators.

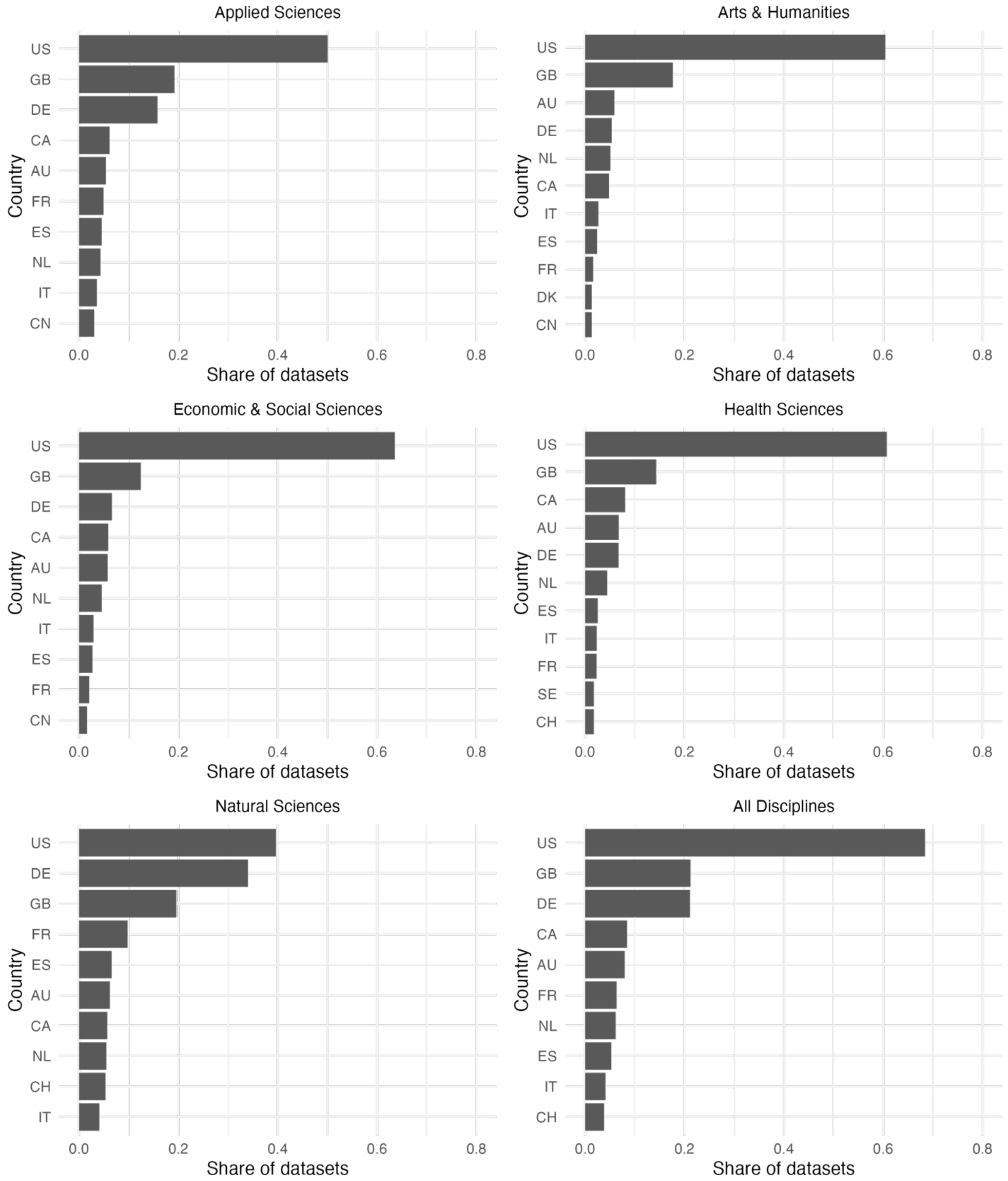

Figure 4. Top 10 dataset producing countries by discipline.

## Dataset citers

The 83,986 works citing datasets are attributed to authors working at 17,072 institutions across 190 countries. The top ten institutions, per discipline and overall, for the authorship of data-citing works, are shown in Figure 5. The top citing institutions differ considerably from the list of those producing datasets, with only one common entry in the overall top ten (University of Washington). The list of overall citers is dominated by high-producing institutions in the Health Sciences discipline, owing to the greater number of data-citing works in this category. At both the institutional and overall country level, we see the entry of other countries into the top ranks, notably China, indicating a higher production of data-citing work utilizing datasets produced elsewhere. Figure 6 shows the top ten countries for authorship of data-citing works.

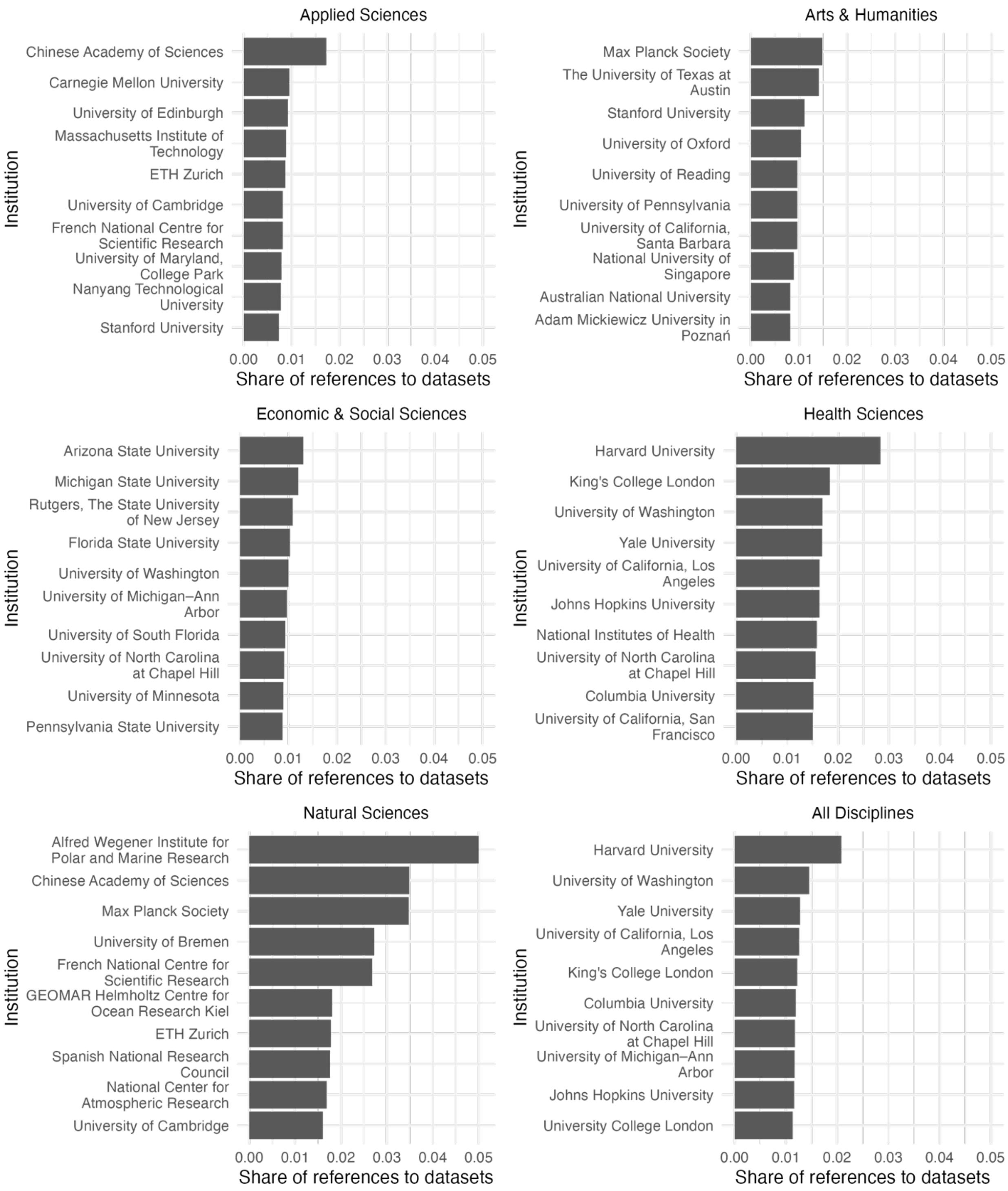

Figure 5. Top 10 dataset citing institutions by discipline.

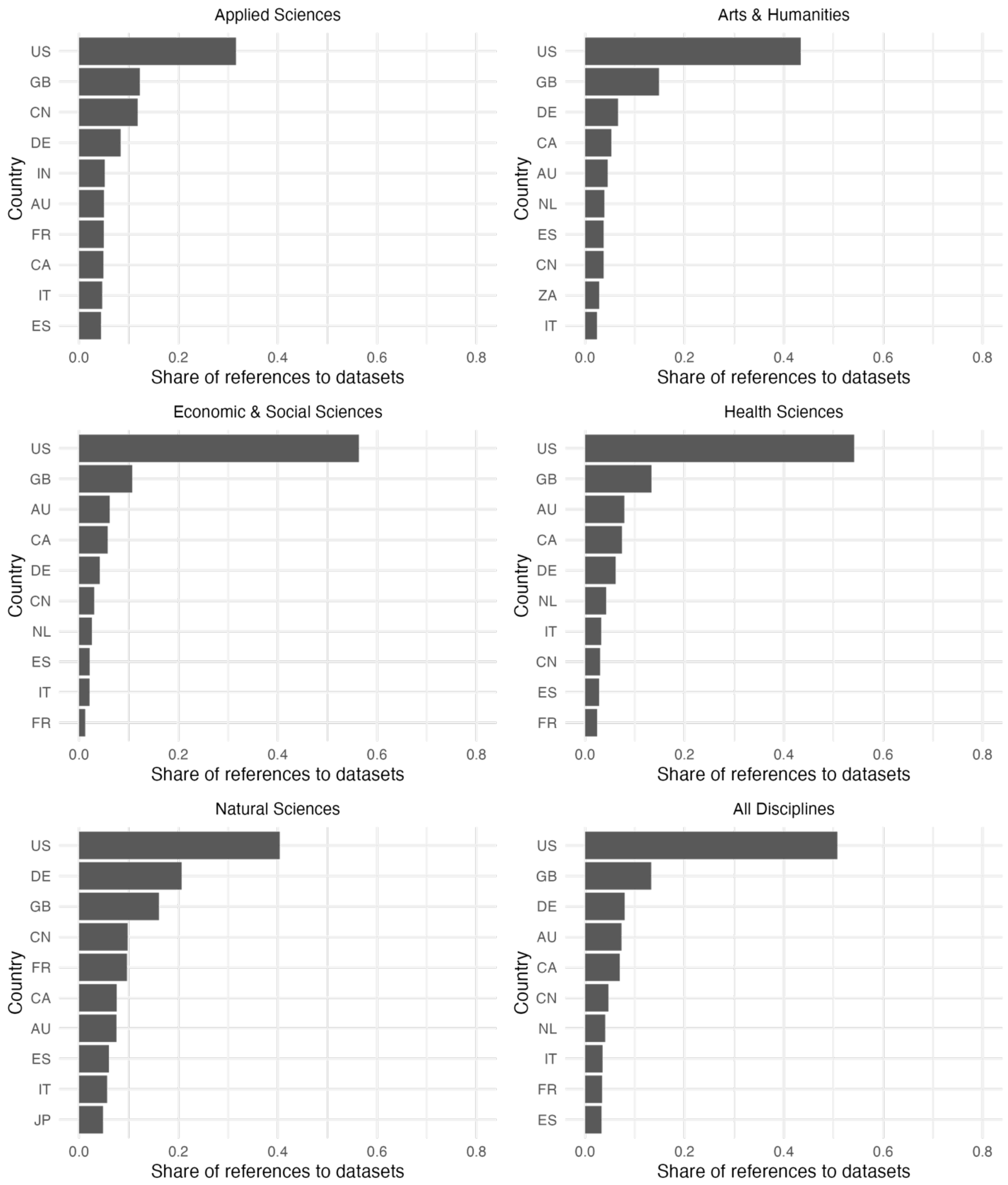

Figure 6. Top 10 dataset citing countries by discipline.

## Relationship between the dataset citers and producers

### Self-citations at the author, institution, and country level

Author matching resulted in 7,224 citations with identified author matches, or 7.86% of all citations, and datasets with self-citations accounted for over 36% of all cited datasets. Self-citations were highly clustered around the less-cited datasets, with much lower proportions in the more-cited datasets, compared to the distribution of overall citations (see Figure 7).

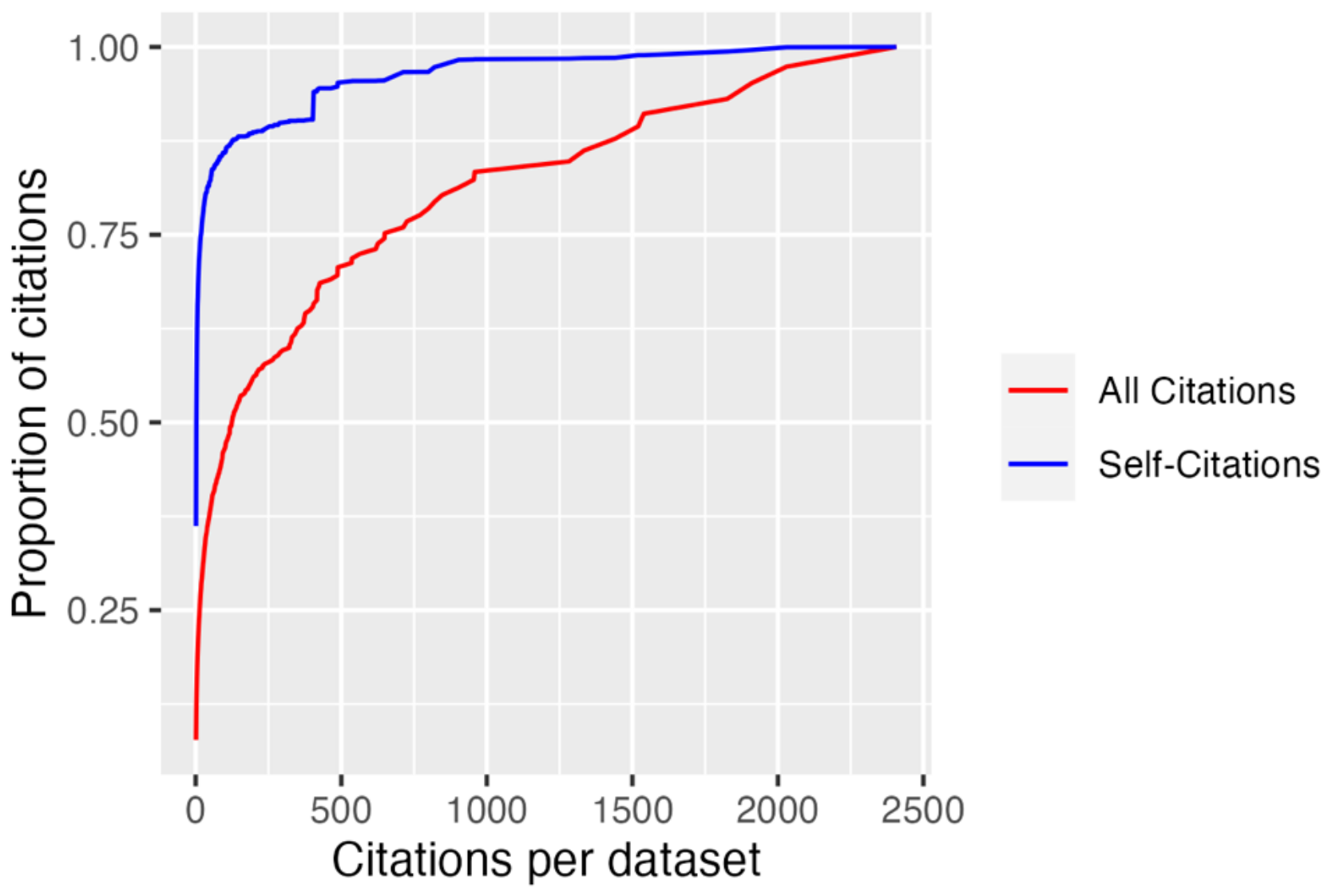

Figure 7. Cumulative proportion of citations to datasets by number of citations

The Natural Sciences show the highest rate of self-citation of datasets by far, at 22%, nearly three times the overall rate. Applied Sciences also show a high rate of self-citation, at nearly 12%. In contrast, while Arts & Humanities had a similar number of citations per dataset to these two disciplines, it has a much lower rate of self-citation, less than 6%. Health Sciences had the lowest rate of self-citation, at 4.6%.

Table 4. Individual self-citation rates by discipline

| | Citations | Citing Works | Distinct Datasets | Self-citation rate |
|---|---|---|---|---|
| Overall | 91,862 | 86,233 | 11,913 | 7.86% |
| Applied Sciences | 7,881 | 7,530 | 2,437 | 11.9% |
| Arts & Humanities | 1,561 | 1,518 | 593 | 6.15% |
| Economic & Social Sciences | 13,725 | 13,043 | 2,653 | 4.92% |
| Health Sciences | 56,537 | 53,513 | 5,134 | 4.93% |
| Natural Sciences | 12,595 | 11,046 | 4,491 | 22.0% |

At the institutional level, the Natural Sciences show the highest citation rate (29.3%). The Applied Sciences (13.7%) and Health Sciences (7.43%) follow. The Arts & Humanities show little growth beyond individual re-use.

Of the 71,189 citations with available author institution data, 7,546, or about 10%, involve citing works and datasets produced at the same institution. Overall, 1,672 institutions have matched citation pairings, though only 896 have more than one such citation. Note that citations may have multiple institutional matches.

All citations with institutional data available for both the citing work and dataset likewise had country data available, and 45,381 of these citations had matching countries between works, a rate of 56.5%. In total, 104 countries had works that cited datasets produced by the same country, with 84 having more than one such citation. At the discipline level, re-use of datasets within the countries where their producers are based ranges from just over 40%, for Applied Sciences, to nearly 60% for Economic & Social Sciences.

Table 5. Institutional and country self-citation rates by discipline

| | Citations | Citing Works | Distinct Datasets | Institution citation rate | Country citation rate |
|---|---|---|---|---|---|
| Overall | 71,189 | 67,863 | 8,023 | 10.6% | 56.5% |
| Applied Sciences | 5,714 | 5,538 | 1,626 | 13.7% | 40.5% |
| Arts & Humanities | 1,082 | 1,064 | 357 | 6.84% | 47.1% |
| Economic & Social Sciences | 10,079 | 9,740 | 1,748 | 6.47% | 59.9% |
| Health Sciences | 45,391 | 43,580 | 3,411 | 7.43% | 57.7% |
| Natural Sciences | 9,200 | 8,207 | 3,183 | 29.3% | 57.8% |

## Cross-institutional data re-use

Figure 8 presents a network of institutions linked through the top 10 edges with the highest weight for each discipline. The edge list is created with only the data citation instances where none of the only cases where none of the citing work institutions matched any of the dataset institutions. Some of the institutions are amongst the top dataset re-users or producers in more than one discipline. We assign them to the discipline in which they have the highest share of citing work overall to avoid duplicating those nodes in the network. In the network, the node size is based on the number of outgoing edges (data re-uses by the institution) and the label size is based on the number of incoming edges (citations received by datasets produced by the institutions). The colors represent the discipline: Applied Sciences (purple), Arts & Humanities (blue), Economic & Social Sciences (fuschia), Health Science (orange), and Natural Sciences (green). The network highlights the role of important data centres, most importantly the Centre for Policy Research in Economic and Social Sciences, and the presence of several Ocean and Marine research data centers in the Natural Sciences. Some universities also stand out as important data providers for other institutions. Also clear in the network is the disciplinary focus of data infrastructures evidenced by the limited overlap between the core data providers in the different disciplines.

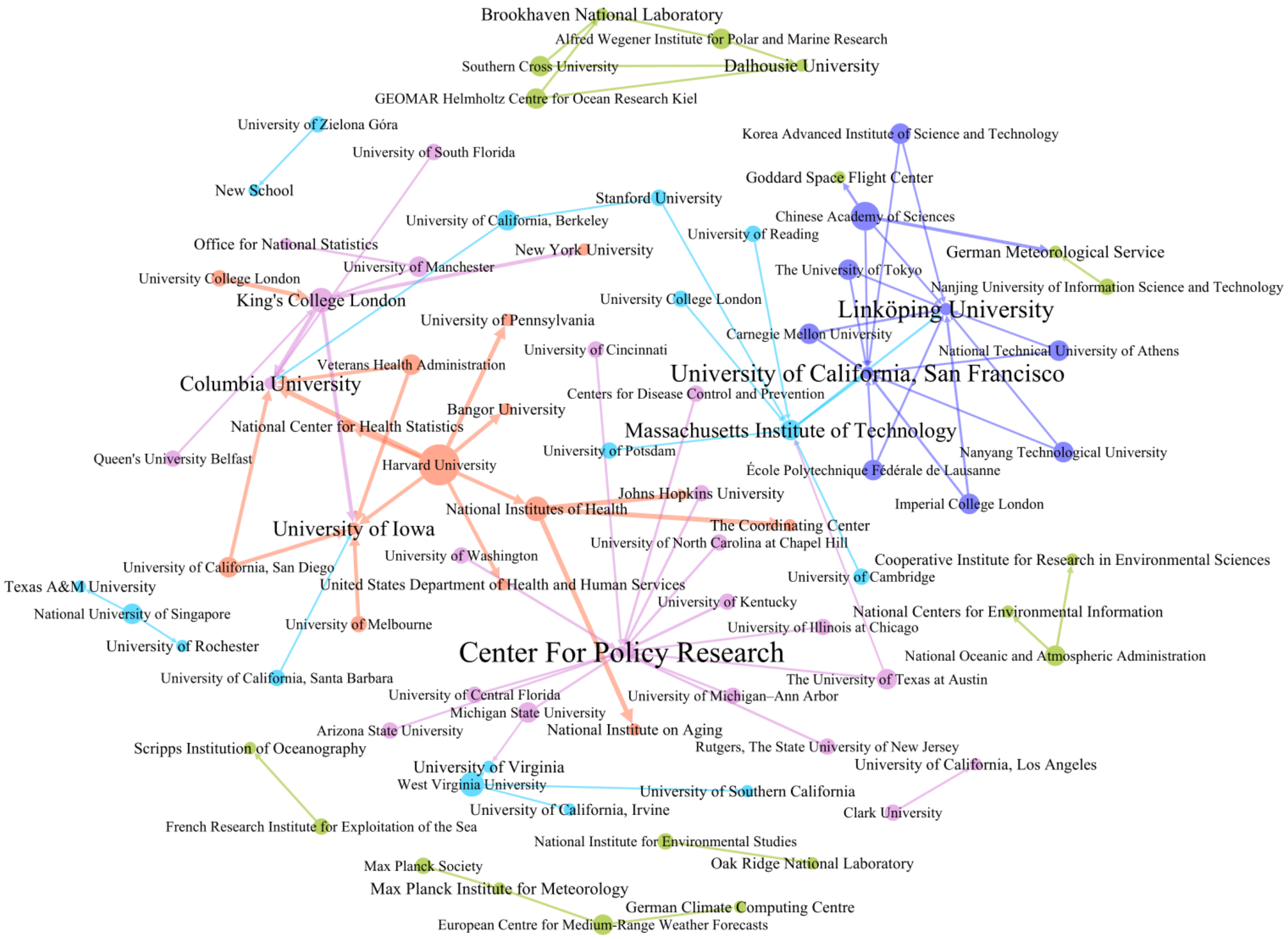

Figure 8. Cross-institutional data re-use network

## Cross-country data re-use

Figure 9 presents the top countries who have the larger share of data imports (re-use of data produced outside of the country) and data exports (datasets cited by other countries). Data exports are heavily concentrated, with the US accounting for more than 50% of the total, and the top 10 main exporters accounting for 88.1% of all exports. In comparison, the data imports are more evenly distributed, with the top 10 data importers accounting for 57.1% of total imports. Notably, there are differences in the list of countries appearing in the two lists. China and Italy are part of the top 10 importers only, and Sweden and Hungary show up in the top ten data exporters.

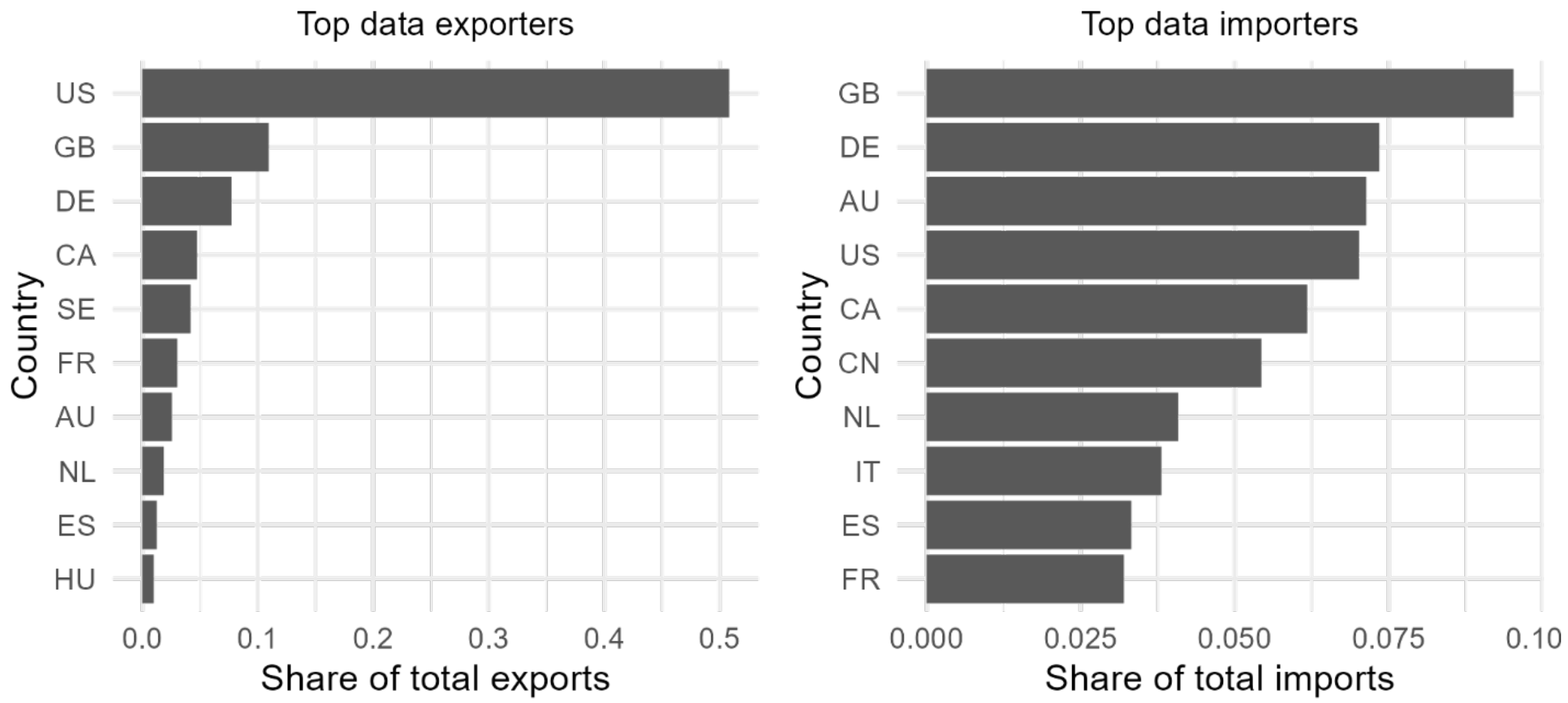

Figure 9. Top 10 countries with the largest share of data exports (left) and imports (right).

**Cross-region data re-use**

At the level of broader geographic regions, Figure 10 clearly shows that researchers in Canada and the United States, and, to a lesser degree, Europe, produce more citations to datasets created by researchers in their own regions, while other regions produce more citations to these two regions than to their own locally-produced data.

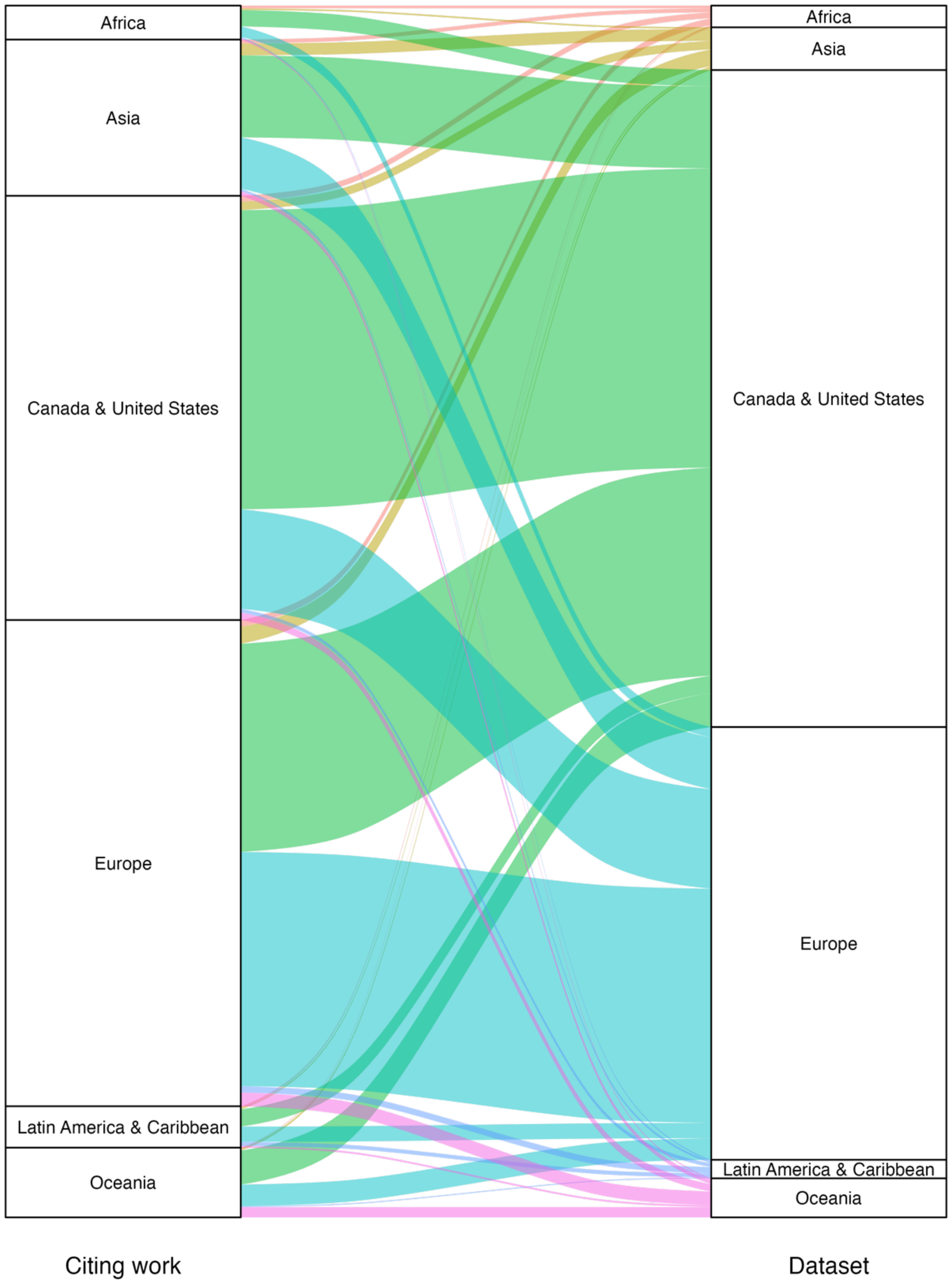

Figure 10. Citation patterns by geographic region

# Discussion

The rate of self-citation to datasets by individual researchers was found to be significantly lower than was identified by Dudek et al. (2019) – 7% versus 75% – but is comparable to the 8% found by Park and Wolfram (2017). Our findings reveal a low rate of both self-citation and institution-matched citations for datasets, suggesting the possibility that authors of works utilizing data are not citing datasets upon initial use, and that formal data citation may be more common when the dataset originates elsewhere. Self-citations were highly clustered around the less-cited datasets, with much lower proportions in the more-cited datasets. This may indicate the limited usability of datasets outside of their original context and the relatively small range of research questions that a dataset can be used to answer (Fan et al., 2023), in line with our hypothesis that shared datasets may be more useful and usable for those who shared them (Pasquetto et al., 2019). Our findings thus call into question the oft repeated argument that sharing data enable further discovery through its re-use. Sharing data may satisfy a policy or mandate and signal transparency but, despite their limitations, our results suggest that data sharing will rarely lead to replication or new knowledge production that can be identified through a formal citation of the shared data.

Institutional self-citations also comprised a relatively small proportion (approximately 10% of citations for which institutional data was available). These findings contrast with Zimmerman's (2007) claim that data sharing and re-use can rely heavily on proximal social interactions, and Jiao and Darch's (2020) findings that data produced within the same institution may be more accessible to researchers. That researchers are not only citing data produced within their own institution could also result from efforts to integrate data infrastructures between institutions or across research fields.

*Disciplinary differences*

Notable disciplinary differences are observed in the frequency of data citations at individual, institutional, and national levels, reflecting how disciplinary data needs, formats, and customs are a major factor in data re-use and citation practices. The variation in data citation rates between disciplines reflects differences in how data is shared, (re-)used, and cited, as well as in the nature of the data itself. In the Health Sciences, the average number of citations per dataset is more than double that within any other discipline. Coupled with the high level of in-country re-use, this would seem to indicate more effective data-sharing regimes, and an emphasis on re-using health data on local populations. The low rate of self-citation in the Health Sciences may reflect norms around citing data creators within the discipline. Paired with the higher average rate of citations in the field, this could indicate a greater ability within the discipline to mobilize data outside the immediate circle of those gathering it. The high rate of institutional citations may be due to the prevalence of health-focused institutions such as the NIH among producers/citers of datasets.

The high institutional citation rate of the Natural Sciences, meanwhile, may reflect an institutional drive to extract maximum value from data produced by costly equipment or complex experimentation, or a high degree of specialization by institutions. Likewise, the high level of intra-country re-use of data within the Social Sciences could be attributed to its population-specific nature. Conversely, one might assume that data in fields within the Applied Sciences, including engineering and computer science, might be seen as less local in nature, facilitating its use in environments further removed from the data's initial production.

The low levels of re-use of data in general within the Arts & Humanities discipline could be attributable to the less data-focused nature of research in these areas, meaning that occasions to re-use any given dataset would be few and far between, or concentrated in specific areas (e.g., digital humanities). The Arts & Humanities low self-citation rates at the institutional level possibly reflects a high degree of specialization of individual researchers and the data they use.

This study further supports past accounts of disciplinary differences in data standards and practices, which influence eventual re-use. Disciplines such as the natural sciences which more often require data management plans may influence the ability of researchers to share it, enabling its re-use. Similarly, funding agencies and organizations that present different requirements across disciplines influence its generation, storage, preservation, and availability (Tenopir et al., 2020). These factors may make the simple difference of what data is available for re-use and citation, and what data is not.

## Limitations

While our study achieves its goal by providing insights into the distribution of data sharing and re-use and the relationship between the data producers and the data citers, some important limitations must be acknowledged. First, our dataset is essentially a convenience sample comprised of the datasets included in OpenAlex. Second, the matching of dataset creators to citing authors remains a challenge, given the complexities involved both in personal naming conventions and data quality. Similarly, the lack of institutional metadata in OpenAlex dataset authorship records is a barrier to investigating this aspect of re-use, as well as in using it to confirm author matches. Third, formal citations are an imperfect measure of data re-use, as datasets may be discussed or mentioned in a publication without actually being used in the research. Conversely, since formal data citation practices are not yet well-established, datasets may be acknowledged through other means in papers, which may not be captured in citation data (Park et al., 2018; Park & Wolfram, 2017). Fourth, our study does not differentiate between the types of datasets used and the context in which they were shared. This may cause an over-representation in our dataset of data generated by large data infrastructures specifically for the purpose of supporting research, whereas datasets shared alongside research papers for the purpose of transparency may be overrepresented in the set of uncited datasets that were excluded from our analysis or concentrated in the long tail of data producers and re-uses.

Further work to complement our findings is needed using different data sources and different methods for identifying data citations and differentiating between different types of datasets (e.g., data produced in the context of a specific research output and shared alongside it, vs. datasets created specifically to support research needs of a research community).

# Conclusion

Exposing the current state of data sharing and re-use by measuring data citation has important implications for data as a scholarly object, the movement towards Open Science, and factors that influence the research landscape, such as policy and funding bodies (Groth et al., 2020). Over 90% of datasets in OpenAlex remain uncited, and within the cited datasets most are cited only once (and often by the creators of the dataset themselves), while the vast majority of citations are made to small proportion of datasets.

Exact matching of names and identifiers, plus a simple approach to name variation matching, identified fewer than 10% of citations as self-citations. However, this rate increased as we looked at less-cited datasets, which comprise most of the overall number of datasets with citations. The citing of datasets by researchers within the same institution occurred at a similarly low rate, when comparing institutional affiliations for first authors, while citations within the same country occurred at a rate of 58%.

Understanding where and how the sharing of data between researchers, institutions, and countries takes place may help to further develop research practices and collaboration. But the effectiveness of such investigations depends strongly upon the available data sources, starting at individual researchers' willingness to cite their use of datasets, through to bibliometric databases' ability to capture and present this information in a useful fashion.

This study may provide insight into which FAIR principles may benefit from concerted efforts made towards advancing them. Data sharing and citation is evidently closely interlinked with the scientific rewards system. Efforts are being increased to link data creators to their data, to better incorporate them into the rewards system and incentive data sharing and re-use (e.g., Mongeon et al., 2017). While willingness on the part of researchers is present and data sharing policies and mandates abound, it remains a challenge to track data re-use and effectively offer credit to researchers, limiting incentive (Tenopir et al., 2020). Monitoring mechanisms on data-sharing platforms and other means of aiding the establishment of data metrics may be useful ways of integrating incentives into Open Science policies. Indeed, what the literature, and the findings of this paper show, is the interplay between data citation practices and their driving forces. Further research should focus on how to successfully advance Open Science using the strong influence of incentives in policy development through the mechanism of infrastructure and technology; that is, swimming with the norms of science and disciplinary idiosyncrasies, not against them.

## Open science practices

The original data for this study was obtained from OpenAlex and is freely available under a Creative Commons CC0 license. Processed data is available on Zenodo using DOI https://doi.org/10.5281/zenodo.7853027.

## Acknowledgements

This research received no specific grant from any funding agency in the public, commercial, or not-for-profit sectors.